# Path-integrated electrostatic energization in laser-driven collisionless cylindrical convergence

D. Pan ,* M. Murakami ,† S. Fujioka, and H. Nagatomo
*Institute of Laser Engineering, The University of Osaka, Suita, Osaka 565-0871, Japan*

(Received 12 July 2026; accepted 29 September 2026; published 15 October 2026)

Collisionless converging ion flows can sustain particle energization over an extended path, but the field channel responsible for the energy gain is not always identifiable. In laser-driven bladed microtube implosion, energetic carbon ions were previously observed alongside a self-generated gigagauss-class axial magnetic field [D. Pan and M. Murakami, Phys. Plasmas **32**, 072107 (2025)], leaving the origin of their energization unresolved. Using two-dimensional particle-in-cell simulations, field-particle power decomposition and trajectory-resolved work accounting show that radial electrostatic work accumulated along the converging path supplies more than 98% of the net in-plane electric work, whereas the azimuthal channel associated with the time-varying magnetic field remains subdominant. The peak local mean ion energy factorizes, to good approximation, into an incident-flow contribution controlled by laser intensity and a geometric enhancement that grows approximately linearly with the inner radius because ions accumulate electrostatic work over a longer convergence path. The path-averaged radial electric field $\langle E_r \rangle$ scales with the peak laser intensity $I_L$ as $\langle E_r \rangle \propto I_L^{0.47}$, favoring a geometry-pinned, comoving charge-separation structure over Debye-limited scaling. The peak local mean carbon-ion energy reaches 0.89 GeV. A smooth-cylinder control exhibits comparable peak energization, showing that blade modulation mainly affects poststagnation ion residence rather than the energy source.

## I. INTRODUCTION

Laser-driven ion acceleration encompasses several established mechanisms [1]. In target-normal sheath acceleration, laser-heated electrons establish a strong electrostatic sheath near the target surface that accelerates ions into vacuum [2–4]. Radiation-pressure acceleration instead transfers laser momentum to an overdense target layer [5], whereas collisionless shock acceleration relies on ion reflection from a propagating electrostatic front [6,7]. In these canonical planar or quasi-planar regimes, ion energization is usually characterized in terms of the local sheath, shock-front, or radiation-pressure dynamics.

Converging geometry is qualitatively different: directed energy is focused into a shrinking stagnation region, so geometry can become an active control of the energy gain. The ion energy can then depend not only on the local field strength or dissipation rate, but also on the distance over which the converging particles remain coupled to the accelerating structure. Hydrodynamic converging shocks are well studied in inertial confinement fusion (ICF) and self-similar compression [8–10]. Less established is the kinetic pathway by which a collisionless, cylindrically converging ion flow converts directed energy into a localized non-directed population. Related multistream stagnation occurs in dense Z pinches, laboratory-astrophysics experiments, and kinetic phases of inertial-fusion implosions [11,12]. This problem is related to broader questions of energy conversion in collisionless shocks, counter-streaming relaxation, and nonthermal particle production [6,7,12–16], but convergence introduces the additional question of how the particle trajectory through a self-consistent field contributes to the final energy gain.

Convergence has also been exploited in laser-driven ion acceleration. Wang *et al.* demonstrated a shaped solid-density target in which radiation pressure drives a converging ion layer, enabling simultaneous acceleration and focusing [17]. The present geometry is physically distinct: in a hollow cylindrical target, ions are launched from the inner wall and converge through the cavity, where they can accumulate electrostatic work over an extended portion of the inward path rather than being accelerated directly by radiation pressure. Microtube implosions have previously been studied primarily as platforms for generating megatesla-class axial magnetic fields [18–21]. The role of cylindrical convergence in the accompanying ion energization, however, has remained unresolved.

In bladed microtube implosion (BMI), blade modulation produces gigagauss-class axial fields without an external seed and is accompanied by carbon ions with energies of several hundred MeV [22]. Their energy source, however, was unresolved: possible candidates included inductive acceleration by the evolving magnetic field, shocklike heating, and electrostatic acceleration along the converging path. The regime is genuinely kinetic. For the initial ion density $n_{i0} = 3 \times 10^{22}\,\mathrm{cm}^{-3}$ and characteristic ion speed $v \sim 0.1c$, where $c$ is the speed of light, the ion-ion Coulomb collision time exceeds the ~100 fs stagnation time by more than an order of magnitude. Sheath acceleration, geometric convergence, electromagnetic-field generation, and phase-space mixing therefore coexist without collisional relaxation.

*Contact author: pan.diya@e4.kepco.co.jp
†Contact author: murakami-m@ile.osaka-u.ac.jp

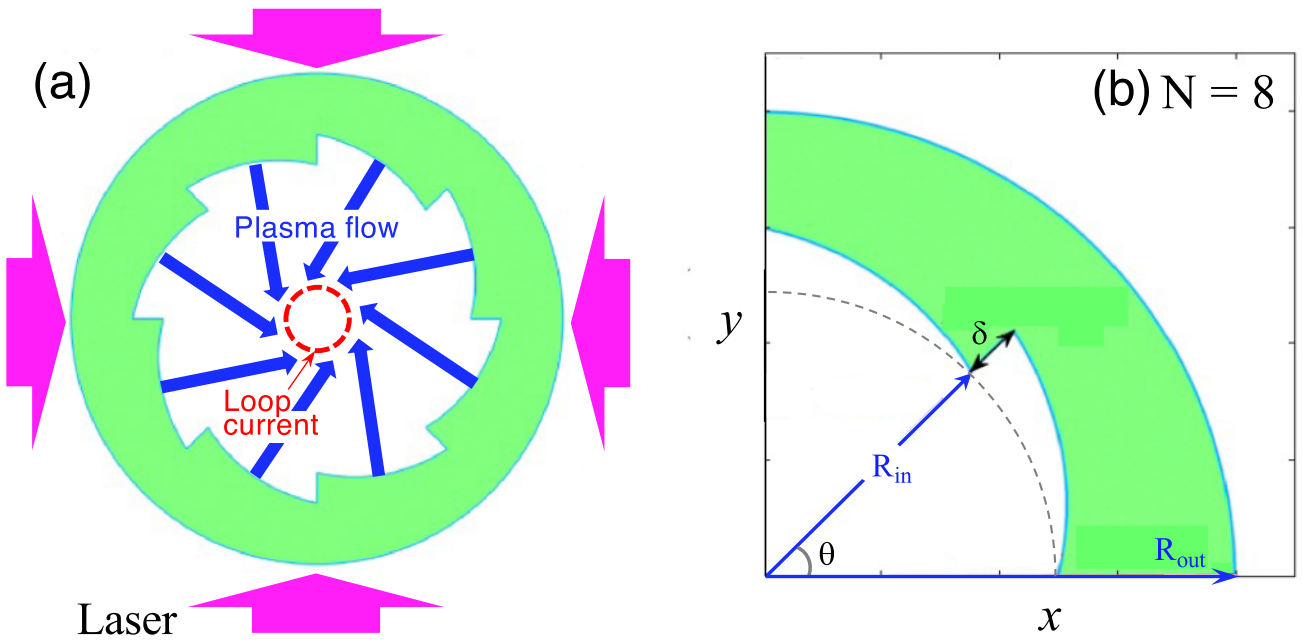

FIG. 1. (a) BMI geometry and four-sided laser drive. (b) Definitions of the inner radius $R_{in}$, outer radius $R_{out}$, blade depth $\delta$, and polar angle $\theta$. Adapted from Ref. [22] under the Creative Commons Attribution (CC BY) license.

Here we identify the dominant ion-energization channel and establish a reduced scaling for the peak ion energy. Ensemble and single-particle diagnostics show that ions gain energy mainly through radial electrostatic work accumulated along the convergence path. The peak energy approximately factorizes into an incident-flow contribution controlled by laser intensity and a radius-dependent geometric enhancement. Smooth-cylinder runs further show that convergence controls the peak energization, whereas blade modulation mainly organizes the poststagnation magnetic environment and ion residence. These results separate energy input, geometric amplification, and poststagnation confinement into approximately separable physical controls. More broadly, they show that ion energization in collisionless converging systems can depend on the particle's history of coupling to a dynamically maintained self-consistent field, rather than being determined solely by the local field at the launch surface.

## II. SIMULATION SETUP

We use the two-dimensional particle-in-cell code EPOCH [23]. Unless stated otherwise, the geometry and numerical settings follow Ref. [22]. The target is a hollow, fully ionized carbon shell with inner radius $R_{in}$, thickness $\Delta R$, and $N = 8$ sinusoidal blade modulations of depth $\delta$ (Fig. 1). The ion density is $n_{i0} = 3 \times 10^{22}\ \mathrm{cm}^{-3}$, with charge-neutralizing electrons. Four linearly polarized laser pulses with wavelength $\lambda = 0.8\ \mu\mathrm{m}$, full width at half maximum (FWHM) 100 fs, and peak time 150 fs enter symmetrically from the $\pm x$ and $\pm y$ Cartesian boundaries; $I_L$ denotes the peak intensity of each pulse. The cell size is $\Delta x = \Delta y = 10$ nm, with 100 ion and 200 electron macroparticles per cell and convolutional perfectly matched layer (CPML) absorbing boundaries.

The baseline case uses $R_{in} = 5\ \mu\mathrm{m}$, $\Delta R = 3\ \mu\mathrm{m}$, $\delta = 1\ \mu\mathrm{m}$, and $I_L = I_0 = 10^{21}\ \mathrm{W/cm^2}$, where $I_0$ is the reference intensity. A 20-run matrix spans $I_L/I_0 = 0.25, 0.5, 1, 2$, and 4 and $R_{in} = 3, 5, 7$, and $10\ \mu\mathrm{m}$ at fixed $\Delta R$. The one-dimensional scans below are cuts through this matrix. A smooth-cylinder control has $\delta = 0$ and otherwise uses the baseline parameters.

We define

$$\mathcal{E}_{peak} = \max_t \langle E_k \rangle_{r<1\ \mu\mathrm{m}}, \tag{1}$$

$$v_{imp} = \max_t |\langle v_r \rangle_{v_r<0,\ \mathrm{shell}}|, \tag{2}$$

$$\alpha = \frac{\mathcal{E}_{peak}}{\mathcal{E}_{imp}}, \qquad \mathcal{E}_{imp} = \frac{1}{2} m_i v_{imp}^2. \tag{3}$$

Here $t$ is time, angle brackets denote particle averages over the region and selection specified by the subscript, and $r$ is the cylindrical radius, $v_r$ is the radial ion velocity, $m_i$ is the carbon-ion mass, and $E_k = (\gamma - 1) m_i c^2$ is the relativistic kinetic energy, with $\gamma$ the Lorentz factor. The diagnostic shell is $R_{in} - 1.5\ \mu\mathrm{m} < r < R_{in} - 0.5\ \mu\mathrm{m}$, centered at $R_{sh} \equiv R_{in} - 1\ \mu\mathrm{m}$.

The quantity $\mathcal{E}_{peak}$ is the peak local mean ion energy, $v_{imp}$ is the incident-flow velocity, and $\mathcal{E}_{imp}$ is the corresponding incident-flow kinetic-energy scale. Their ratio $\alpha$ is the convergence enhancement factor. It is an operational measure rather than a universal property of cylindrical convergence: $\alpha = 1$ would correspond to no enhancement of $\mathcal{E}_{peak}$ relative to the incident-flow energy scale, while $\alpha - 1$ roughly measures the additional convergence-associated energization in units of $\mathcal{E}_{imp}$. As shown in Sec. V, this additional enhancement is associated predominantly with electrostatic work accumulated along the inward trajectory.

Because the central distributions are non-Maxwellian, $\mathcal{E}_{peak}$ is an effective local energy rather than a thermodynamic temperature. The shell samples inward-moving ions before they enter the central multistream region. Ions continue to accelerate at smaller radii, so $v_{imp}$ is not the largest local velocity. At $v_{imp} \sim 0.1c$, the nonrelativistic expression for $\mathcal{E}_{imp}$ introduces an error below 1%.

The blade surface is recessed outward from $R_{in}$, so the diagnostic shell lies inside the vacuum cavity and the same definition applies to the smooth and bladed targets. The absolute value of $\alpha$ depends weakly on the shell location, but its parameter trends are robust. Doubling the macroparticle number changes $\mathcal{E}_{peak}$ by 0.1%; halving the cell size changes it by −3.8%, while $v_{imp}$ changes by less than 1% in both tests.

## III. GLOBAL IMPLOSION DYNAMICS

Figure 2 summarizes the baseline evolution. Laser-heated electrons establish sheath fields that accelerate carbon ions inward, reaching $\langle v_r \rangle \simeq -0.15c$ near $t = 220$ fs. As the flow converges at $t \simeq 230$–250 fs, the high-energy population grows rapidly; the mean energy of the upper 10% reaches ~60 MeV on the global average, while the local central energy is substantially larger.

Over the same interval, the coherent inflow weakens and the radial velocity spread rises. Defining the radial velocity spread as $v_{r,\mathrm{rms}} \equiv [\langle (v_r - \langle v_r \rangle)^2 \rangle]^{1/2}$, the ratio $|\langle v_r \rangle|/v_{r,\mathrm{rms}}$ falls below unity within ~80 fs, much faster than the ion-ion Coulomb collision time $\tau_{ii} \sim$ several ps. This ratio alone does not prove thermalization because ballistic axis crossing and orbital motion also increase the second moment. The density increase is moderate, with no global signature of a mature strong shock, and the axial magnetic field reaches several $10^2$ kT near the ion-energy maximum. The temporal coincidence of these changes motivates the phase-space and work diagnostics below.

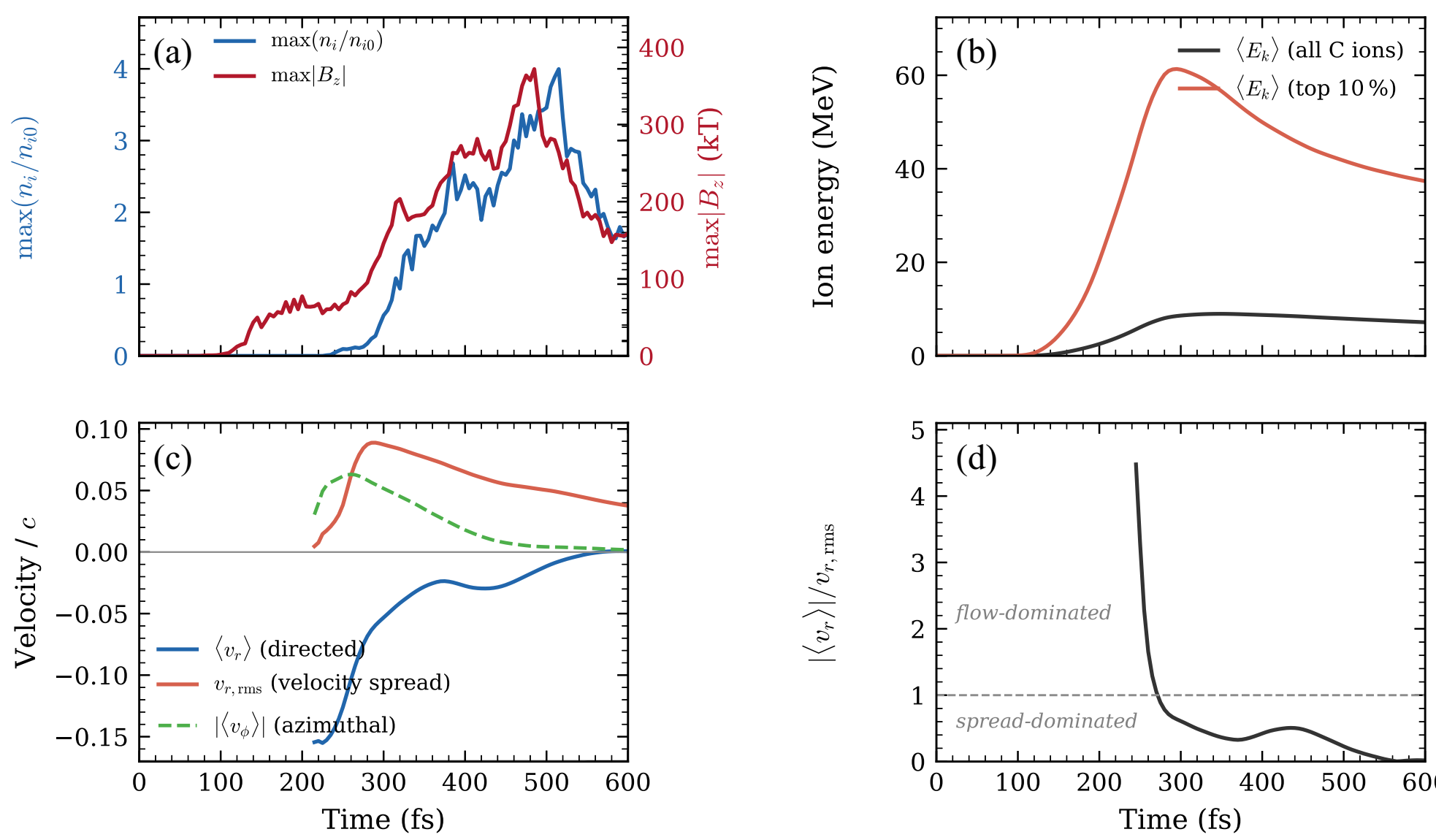

FIG. 2. Baseline evolution. (a) Maximum ion density, normalized to $n_{i0}$, and maximum magnitude of the axial magnetic-field component $B_z$. (b) Mean ion energy and mean energy of the upper 10%. (c) Mean radial and azimuthal velocities and radial velocity spread in $r < 2\,\mu\mathrm{m}$. (d) Ratio of coherent radial flow to radial velocity spread.

## IV. KINETIC MECHANISM OF ION ENERGIZATION

### A. Collisionless stagnation

Figure 3 shows the radial phase space $(r, v_r)$ colored by the magnitude of the azimuthal ion velocity $|v_\phi|$, where $\phi$

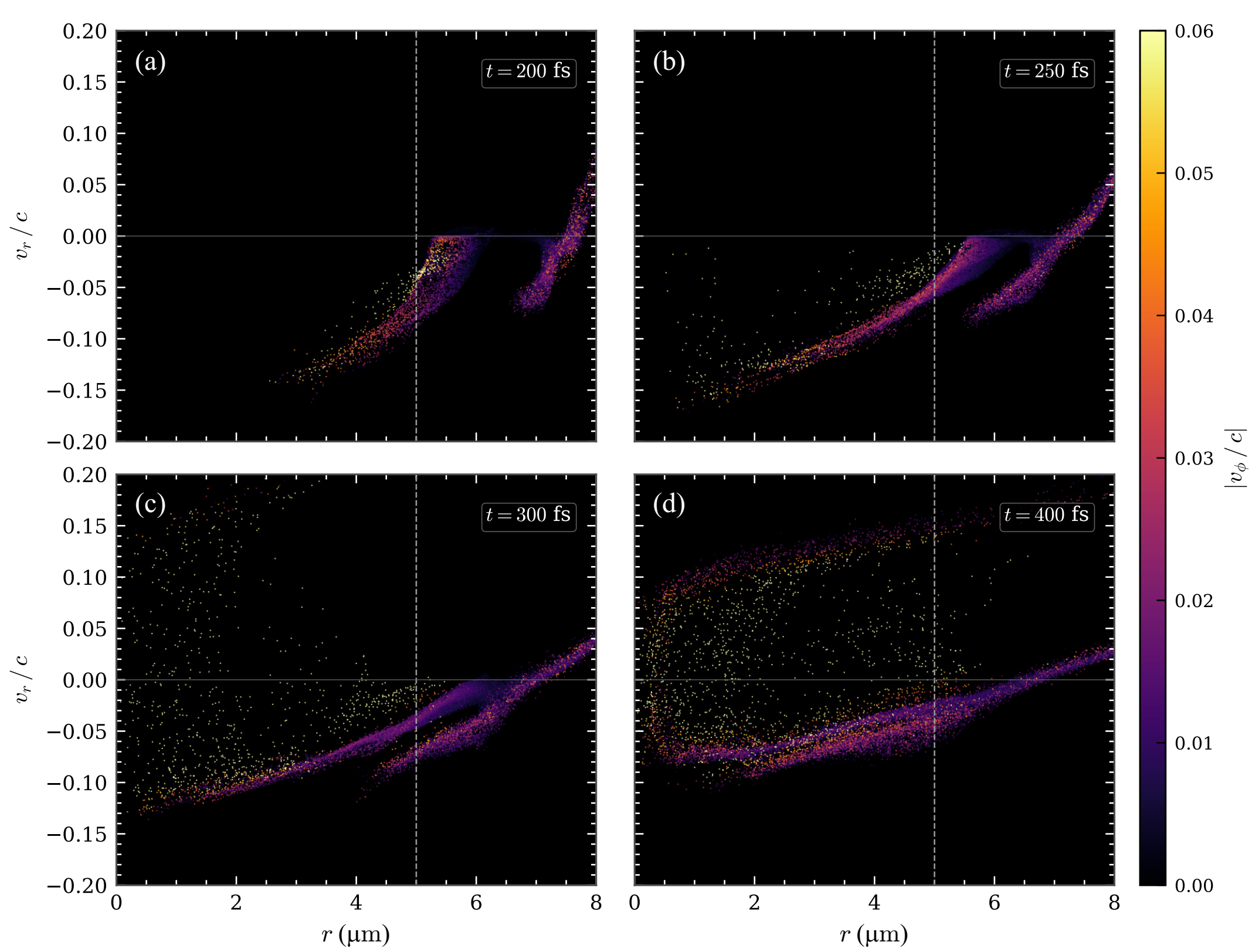

FIG. 3. Carbon-ion phase space $(r, v_r/c)$ colored by $|v_\phi/c|$. (a) Inward beam; (b) arrival at the axis; (c) coexisting inward and outward branches; (d) dispersed poststagnation phase space. The dashed line marks the initial $R_{\mathrm{in}}$.

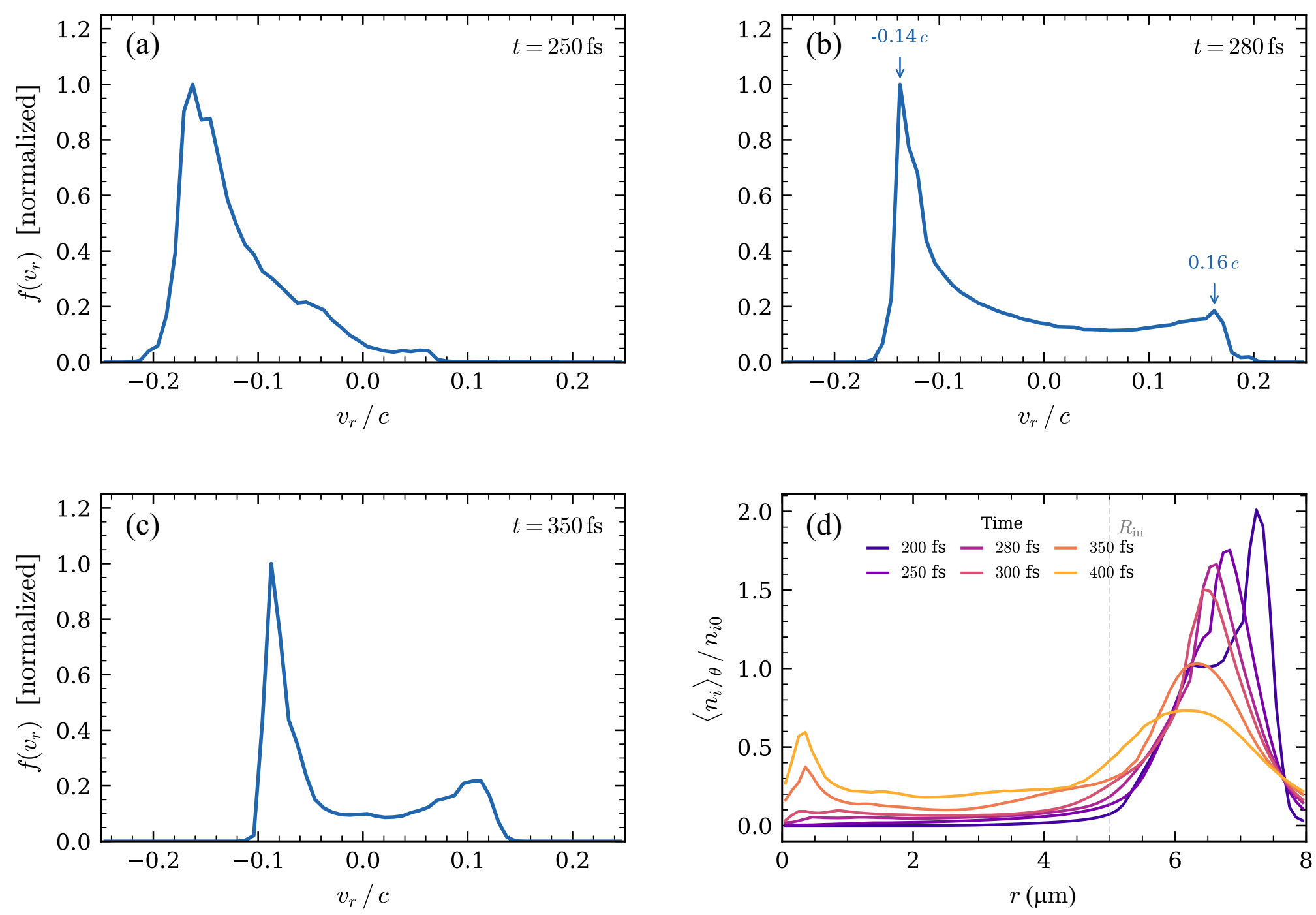

FIG. 4. (a)–(c) Radial velocity distributions $f(v_r)$ in $r < 1\,\mu$m, evolving from an inward beam through axis crossing to a broad non-Maxwellian state. (d) Azimuthally averaged ion-density profiles.

is the azimuthal coordinate. At $t = 200$ fs, ions near the inner wall form a coherent inward beam with little azimuthal motion. By $t = 250$ fs the leading edge reaches the axis, and finite $|v_\phi|$ appears at small radius. At $t = 300$ fs, inward- and outward-moving branches coexist in the central region, showing collisionless axis crossing rather than a single downstream population. Large $|v_\phi|$ near $v_r \simeq 0$ reflects finite-impact-parameter motion near closest approach (the Larmor-hole geometry) [19,21]. By $t = 400$ fs, the initial beam has largely dispersed, although the phase space remains nonthermal.

The central radial velocity distribution gives a complementary view [Figs. 4(a)–4(c)]. It evolves from a peak near $-0.15c$ to inward and outward branches near $-0.14c$ and $+0.16c$, and then to a broad flat-topped distribution. A double peak alone is not evidence of heating: ballistic passage through the axis produces the same polar-coordinate signature (see the Appendix). The more informative features are the subsequent departure from this ballistic form and the absence of a sharp propagating density jump [Fig. 4(d)]. The central density rises by only a few times its initial value and remains spatially broad, rather than forming a coherent discontinuity moving through the incoming plasma. Together these observations indicate collisionless multistream stagnation and incomplete velocity-space redistribution, not mature hydrodynamic strong-shock heating.

### B. Radial field and power decomposition

During acceleration, an inward radial electric field of several TV/m extends from the inner wall toward the stagnation region [Fig. 5(b)]. Unlike a wall-localized Debye sheath, the field spans most of the convergence path. It is consistent with a dynamic charge-separation structure sustained by laser-heated electrons moving ahead of the heavier ions and continually rebuilding the field near the advancing ion front. After stagnation, the field reverses near the axis as an electron-rich core and ion-rich shell form. The ion and electron profiles nevertheless remain close to quasineutral on the plotted scale [Fig. 5(c)], excluding an unscreened non-neutral core as the principal energy reservoir. The relevant electrostatic potential is therefore distributed along the converging path rather than stored predominantly at the axis.

For carbon ions in $r < 3\,\mu$m, the instantaneous in-plane electric powers per ion are

$$P_r = ZeE_r v_r, \qquad P_\phi = ZeE_\phi v_\phi, \tag{4}$$

where $Z = 6$ is the carbon charge state, $e$ is the elementary charge, and $E_r$ and $E_\phi$ are the radial and azimuthal electric-field components. The radial power $P_r$ measures work by radial sheath and space-charge fields along the convergence path. The azimuthal power $P_\phi$ bounds the inductive channel associated with the evolving axial magnetic field, together with any nonaxisymmetric electrostatic work introduced by the blade geometry. During energization, $\langle P_r \rangle$ is strongly positive, whereas $\langle P_\phi \rangle$ remains near zero [Fig. 6(a)]. The cumulative radial channel supplies more than 98% of the net in-plane electric work [Fig. 6(b)].

Two checks show that this coordinate decomposition faithfully isolates the electrostatic contribution. First, the azimuthal average of the radial component of any divergence-free (solenoidal) field vanishes by the divergence theorem, so the coherent radial field is electrostatic. Second, at fixed radius,

$$\langle E_r v_r \rangle_\phi = \langle E_r \rangle_\phi \langle v_r \rangle_\phi + \langle \delta E_r\, \delta v_r \rangle_\phi, \tag{5}$$

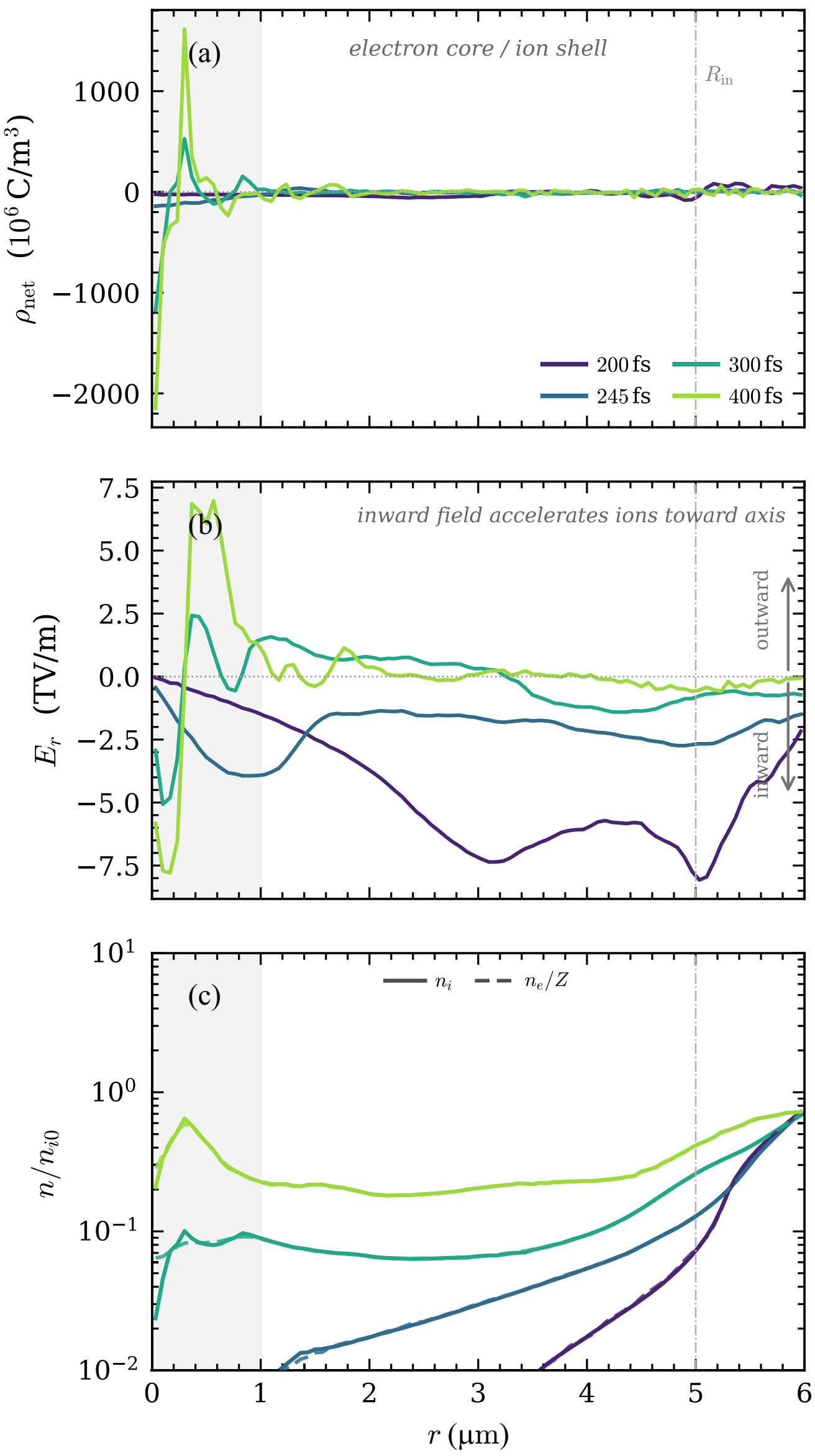

FIG. 5. Azimuthally averaged radial profiles at four times. (a) Net charge density $\rho_{\rm net} \equiv e(Zn_i - n_e)$, where $Z = 6$ is the carbon charge state, $e$ is the elementary charge, and $n_i$ and $n_e$ are the ion and electron number densities. (b) Radial electric field $E_r$. (c) $n_i$ and $n_e/Z$, normalized to the initial ion density $n_{i0}$. The shaded region is $r < 1$ μm; the dash-dotted line marks $R_{\rm in}$.

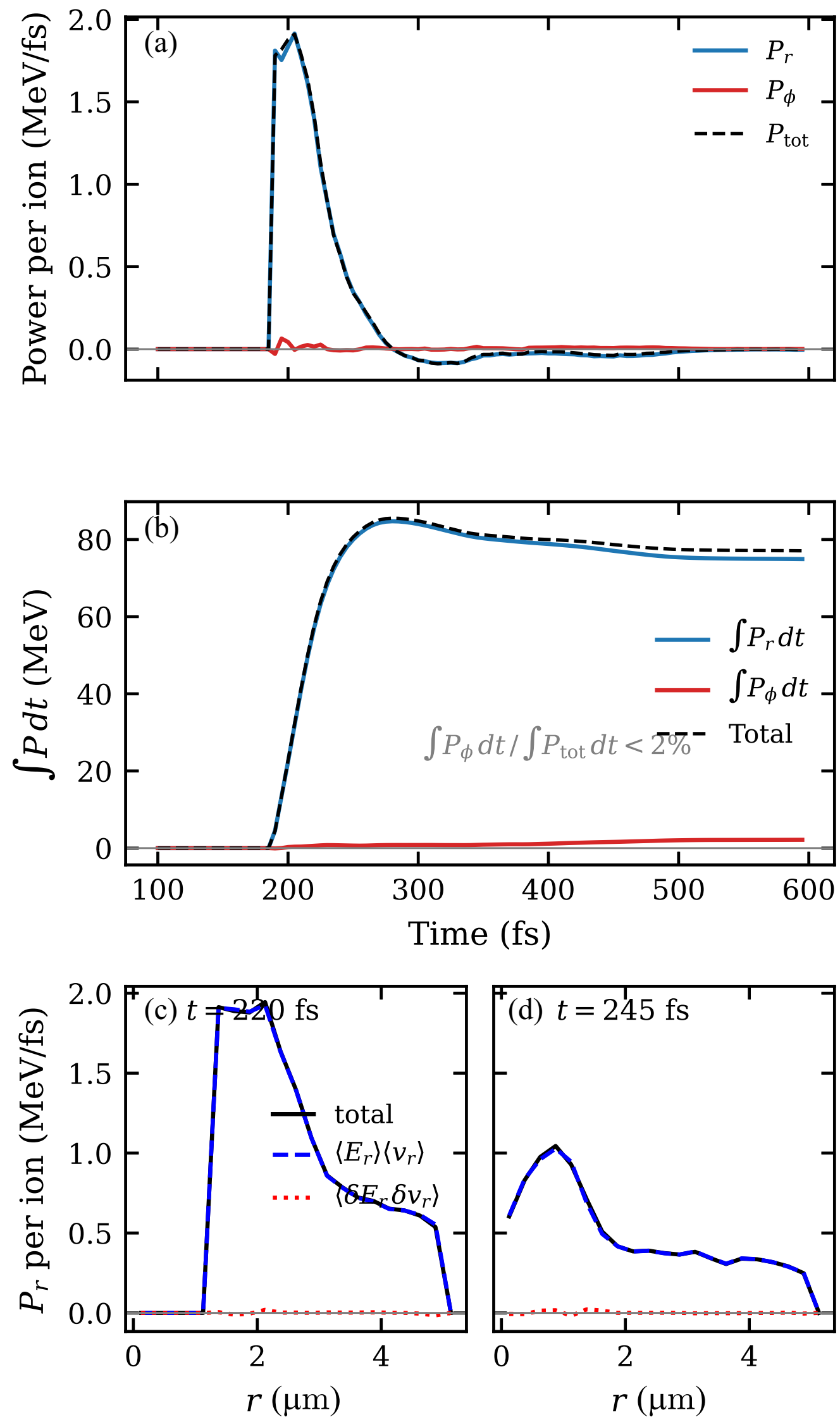

FIG. 6. Field-particle power decomposition. (a) Mean radial and azimuthal power per ion in $r < 3$ μm. (b) Cumulative radial and azimuthal work, $\int P_r\,dt$ and $\int P_\phi\,dt$. (c), (d) Radial power split into coherent and fluctuating parts at 220 and 245 fs.

where $\langle\cdot\rangle_\phi$ denotes an azimuthal average, $\delta E_r \equiv E_r - \langle E_r\rangle_\phi$, and $\delta v_r \equiv v_r - \langle v_r\rangle_\phi$. This decomposition shows that the coherent term supplies at least 98% of the radial work, with the fluctuation correlation remaining within approximately ±1%. The peak-power location follows the ion front inward [Figs. 6(c) and 6(d)], directly visualizing path accumulation. The solenoidal component is nonnegligible in the root-mean-square (rms) field because of transmitted laser light and small-scale structure, but it performs negligible net work on the ions. This distinction is important: a large magnetic-field amplitude or a large rms electromagnetic fluctuation does not by itself imply a correspondingly large field-particle energy transfer.

These results separate energy input from redistribution. Ions gain energy through radial electrostatic work during convergence; the central multistream dynamics then convert the ordered radial flow into a broad, anisotropic velocity distribution. Mechanisms requiring substantial azimuthal electric work, including inductive or betatron-like acceleration by the self-generated axial magnetic field $B_z$, do not control the peak-energy budget.

### C. Energy localization and redistribution

The acquired energy is concentrated near the axis (Fig. 7). The radial maximum enters $r < 1$ μm near $t \simeq 245$ fs and reaches ∼200 MeV before decreasing as the ions disperse.

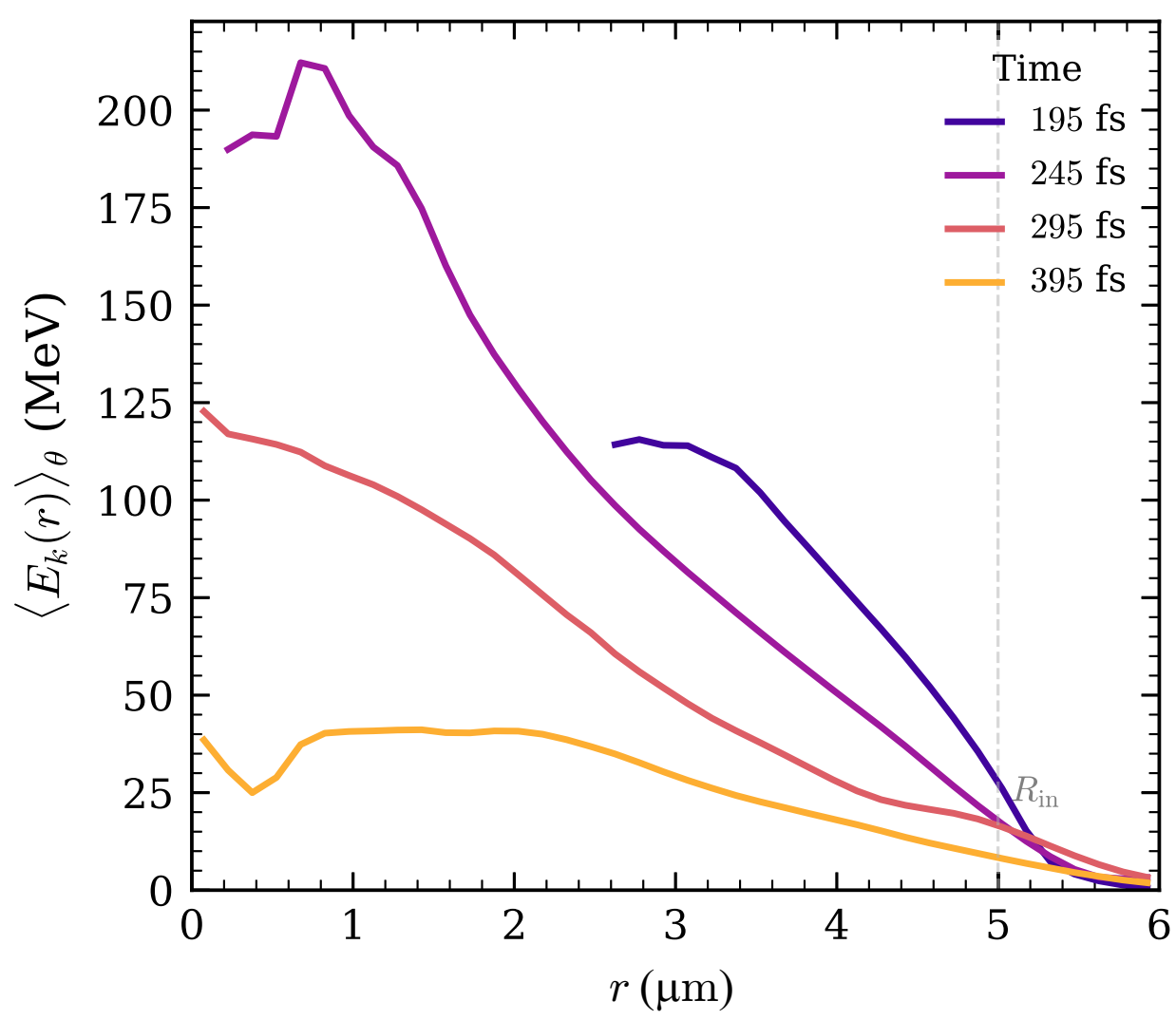

FIG. 7. Radial mean ion-energy profiles. The shaded region is $r < 1$ μm; the dashed line marks $R_{\rm in}$.

Within $r < 2$ μm, we decompose the in-plane ion kinetic energy into directed-radial, velocity-spread, and rotational components:

$$K_r^{\rm dir} = \tfrac{1}{2} m_i \langle v_r\rangle^2, \tag{6}$$

$$K_{\rm spr} = \tfrac{1}{2} m_i [{\rm Var}(v_r) + {\rm Var}(v_\phi)], \tag{7}$$

$$K_\phi^{\rm rot} = \tfrac{1}{2} m_i \langle v_\phi\rangle^2, \tag{8}$$

where ${\rm Var}(v) \equiv \langle v^2\rangle - \langle v\rangle^2$. The fractions in Fig. 8 are normalized to the total in-plane energy, $K_{\rm plane} = K_r^{\rm dir} + K_{\rm spr} + K_\phi^{\rm rot}$. The directed-radial fraction falls from ∼60% to below 5% by $t \simeq 400$ fs, while the velocity-spread contribution rises. The rotational fraction peaks at approximately 17% near stagnation and then decays. This does not imply substantial azimuthal energization: the geometry can redirect radially acquired energy into azimuthal motion without performing appreciable additional azimuthal work.

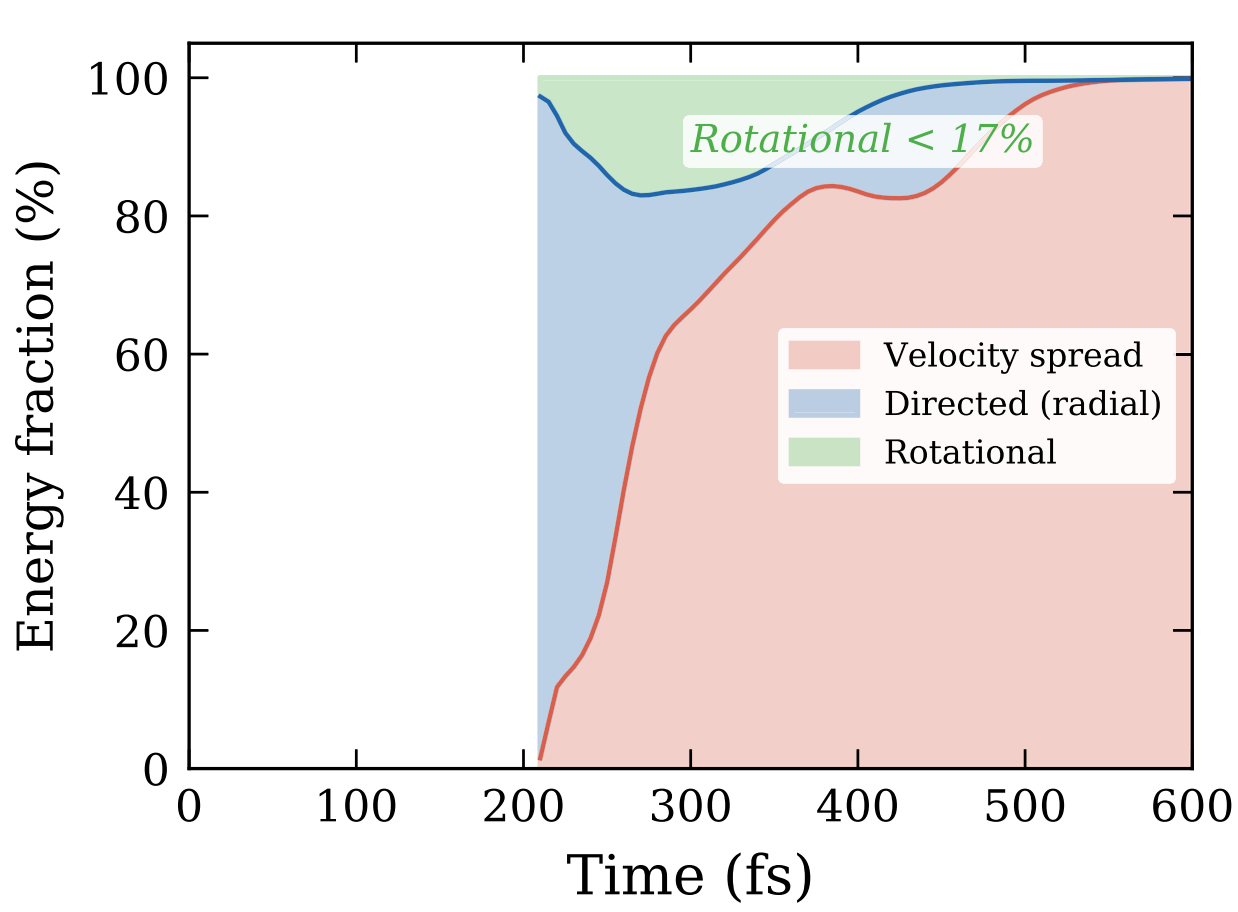

FIG. 8. Energy fractions in $r < 2$ μm: directed radial, velocity-spread, and rotational components.

The redistribution is collisionless rather than collisional: the velocity distribution remains non-Maxwellian and the process time is far shorter than the ion-ion collision time $\tau_{ii}$, so isotropic thermal equilibrium is not reached. Consistently, the velocity-spread anisotropy $T_{rr}/T_{\phi\phi}$, with $T_{rr} \equiv \frac{1}{2} m_i {\rm Var}(v_r)$ and $T_{\phi\phi} \equiv \frac{1}{2} m_i {\rm Var}(v_\phi)$, remains well above unity near stagnation; because ordered ballistic motion also contributes to this ratio, it is not interpreted as a thermodynamic temperature ratio.

## V. PARAMETRIC SCALING AND REDUCED MODEL

### A. Intensity and radius scans

Across the 20-run matrix, laser intensity and inner radius play distinct roles [Fig. 9]. At fixed $R_{\rm in} = 5$ μm,

$$\mathcal{E}_{\rm peak} \propto I_L^{0.41}, \qquad \mathcal{E}_{\rm imp} \propto I_L^{0.37}, \qquad v_{\rm imp} \propto I_L^{0.18}. \tag{9}$$

The velocity exponent is slightly below the ponderomotive expectation of 0.25, plausibly because the absorption fraction varies with intensity. Since $\mathcal{E}_{\rm peak}$ and $\mathcal{E}_{\rm imp}$ scale similarly, $\alpha$ remains nearly constant at each radius. At $R_{\rm in} = 5$ μm, $\alpha = 3.12 \pm 0.13$ (coefficient of variation 4.2%). A weak systematic increase with intensity appears at the larger radii and is quantified below. The radial-work fraction exceeds 98% throughout the intensity scan.

At fixed $I_L = I_0$, increasing $R_{\rm in}$ from 3 to 10 μm raises $\mathcal{E}_{\rm peak}$ from 115 to 488 MeV. In contrast, $v_{\rm imp} = 0.107 \pm 0.006c$, with only a weak ∼15% systematic increase across the radius range. This is consistent with $v_{\rm imp}$ being set mainly by the local sheath condition rather than the global cavity size. Consequently, $\alpha$ rises from 2.16 to 6.68, approximately linearly with radius. The radial-work fraction remains above 97% for all radii.

### B. Trajectory-resolved energy accounting

A low-density, uniquely tagged tracer population, identical in charge and mass to the bulk carbon and seeded at $10^{-3}$ of its density, provides a Lagrangian check. For each of ∼1900 ions, we integrate $\int Ze\,\mathbf{E}\cdot\mathbf{v}\,dt = \int (P_r + P_\phi)\,dt$ along the resolved trajectory, where $\mathbf{E}$ is the local electric field and $\mathbf{v}$ is the tracer-ion velocity, and compare the result with the actual kinetic-energy gain $\Delta E_k$.

Energy grows progressively during inward motion and peaks near the axis rather than at the launch surface [Fig. 10(a)]. The integrated work reproduces $\Delta E_k$ from ∼10 to above 300 MeV with a regression slope of 1.01 and a median relative closure error of 8% [Fig. 10(b)]. The residual scatter is symmetric about the identity line, is consistent with the finite spatiotemporal resolution of the postprocessed fields, and shows no systematic signature of a missing work channel. The radial contribution has a median value of 97% [Fig. 10(c)], independently confirming the ensemble decomposition.

The trajectories also reveal temporal gating. The accelerating field is strongest during approximately 200–260 fs. Ions arriving early therefore traverse the path while the field is near its maximum, whereas late-arriving ions sample a decaying field. Individual energies therefore depend on both path length and arrival time. The parameter scaling below concerns the systematic path-length contribution to the peak population.

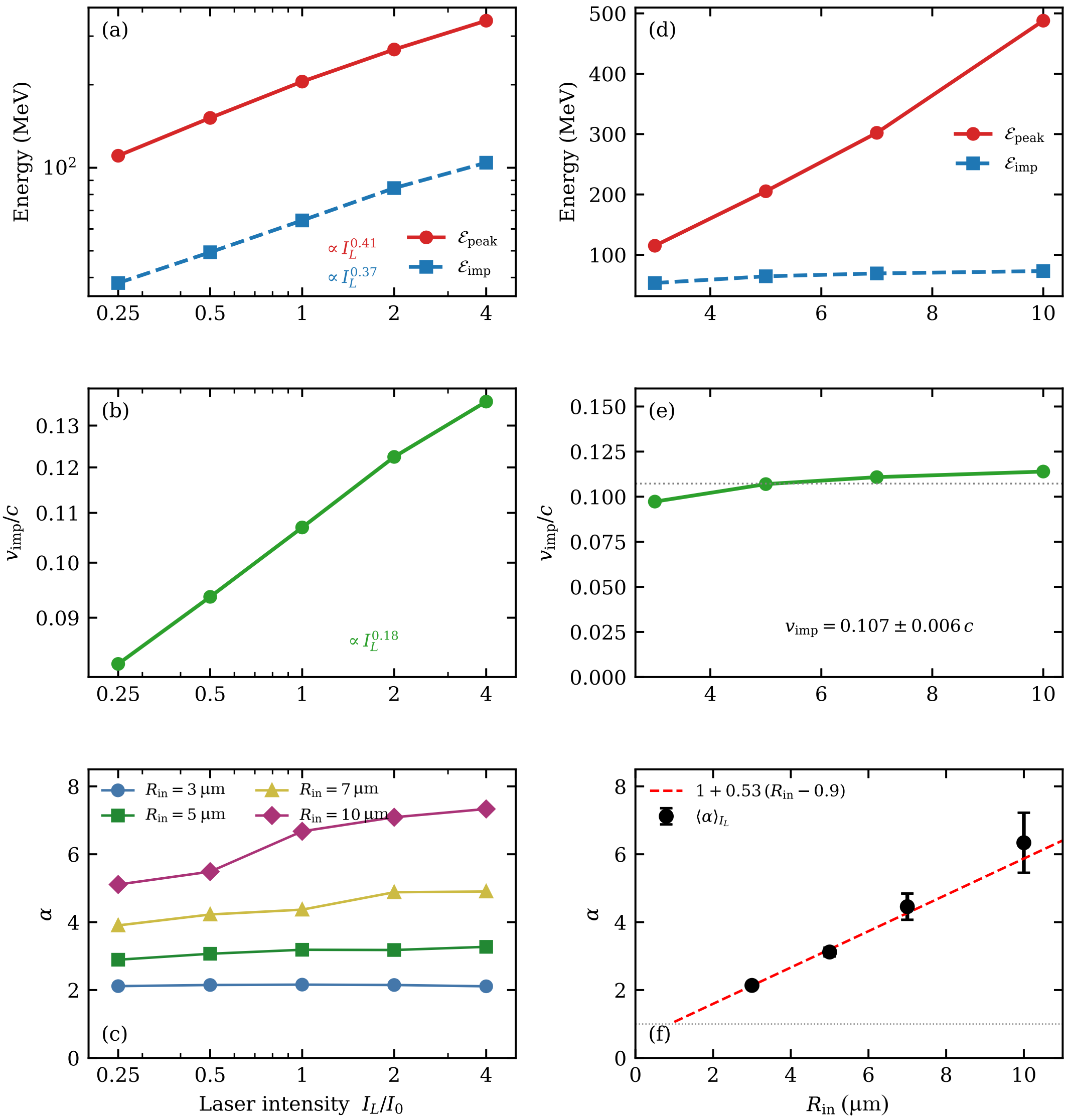

FIG. 9. Parameter scans. Left: intensity dependence of (a) $\mathcal{E}_{\text{peak}}$ and $\mathcal{E}_{\text{imp}}$, (b) $v_{\text{imp}}$, and (c) $\alpha$. Right: radius dependence of the same quantities [(d)–(f)]. In (f), error bars show the standard deviation over intensity and the line is Eq. (14).

### C. Reduced model and field scaling

In the ponderomotive regime, the hot-electron temperature $T_e$ scales as $T_e \propto (I_L\lambda^2)^{1/2}$ [24]. With the ion sound speed $c_s = \sqrt{ZT_e/m_i}$ and $v_{\text{imp}} \sim c_s$,

$$\mathcal{E}_{\text{imp}} = \tfrac{1}{2} m_i v_{\text{imp}}^2 \propto T_e. \tag{10}$$

The characteristic path-averaged radial field may be written

$$\langle E_r \rangle \sim \frac{T_e}{e\lambda_s}, \tag{11}$$

where $\lambda_s$ is the effective scale length of the dynamic charge-separation structure. The field structure comoves with the ion front [Figs. 6(c) and 6(d)], and its amplitude reflects both the local hot-electron pressure and the dynamic charge-separation scale, while increasing $R_{\text{in}}$ mainly lengthens the acceleration path. This is the physical basis of the approximate separability: intensity controls the incident-flow and field scales, whereas radius controls the distance over which ions sample the field before reaching the stagnation core.

To measure this field scale consistently across the parameter matrix, let $r_c^{\text{obs}} \simeq 1$ μm denote the stagnation-core radius measured directly from the simulations. We then define the acceleration-phase snapshot work proxy

$$W_{\text{snap}} = \int_{r_c^{\text{obs}}}^{R_{\text{sh}}} Ze\,|\langle E_r \rangle_\phi|\,dr. \tag{12}$$

The corresponding snapshot path length and path-averaged field are

$$L_{\text{snap}} \equiv R_{\text{sh}} - r_c^{\text{obs}}, \qquad \langle E_r \rangle_{\text{snap}} \equiv \frac{W_{\text{snap}}}{ZeL_{\text{snap}}}. \tag{13}$$

This operational definition does not assume a spatially uniform field. The upper limit $R_{\text{sh}}$ is the center of the shell used to diagnose $v_{\rm imp}$; thus $W_{\rm snap}$ samples radial acceleration interior to the location at which $\mathcal{E}_{\rm imp}$ is defined.

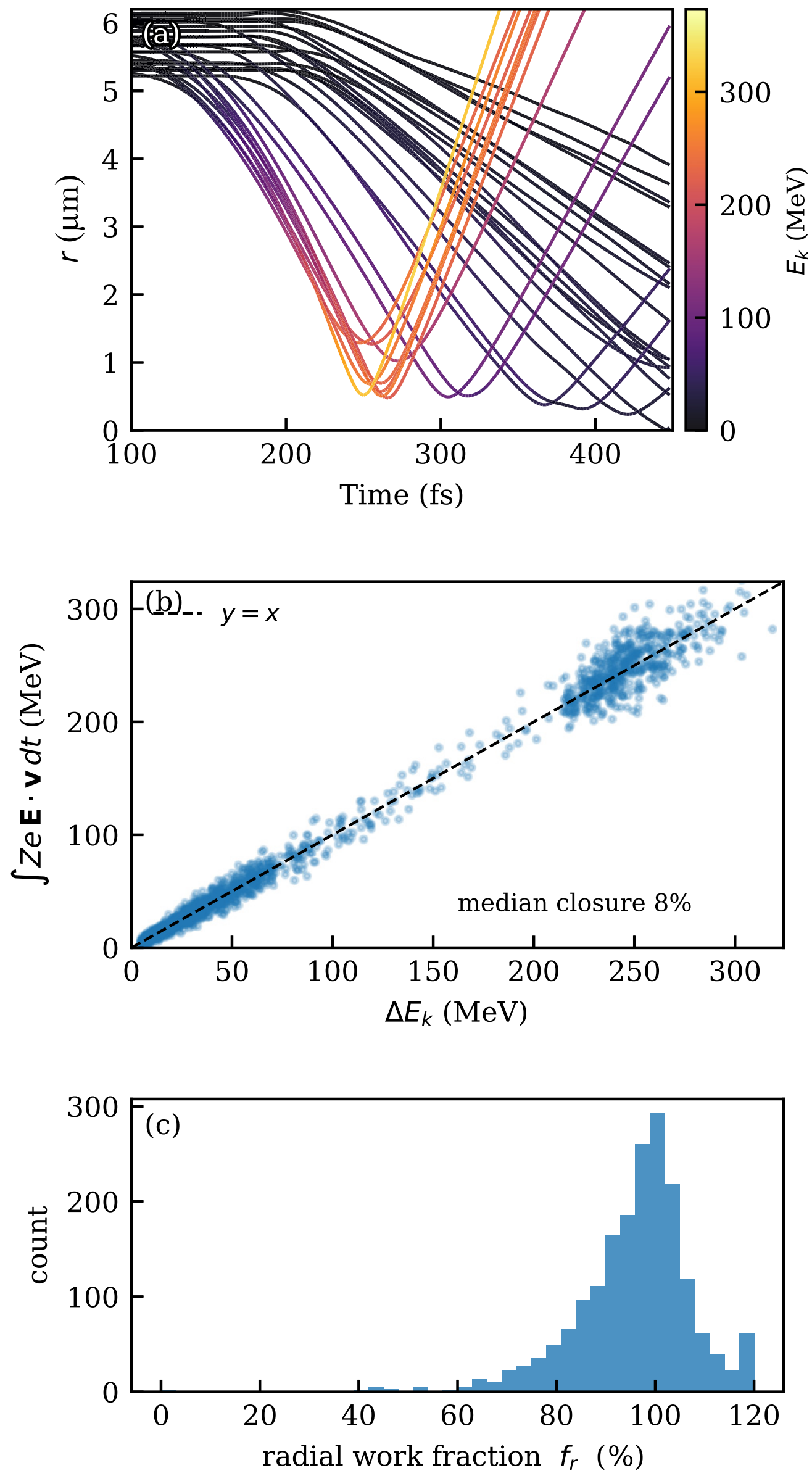

FIG. 10. Tracer-ion energy accounting. (a) Representative radial trajectories colored by kinetic energy. (b) Integrated electric work versus actual energy gain. (c) Distribution of the radial work fraction $f_r \equiv \int P_r\,dt / \int (P_r + P_\phi)\,dt$.

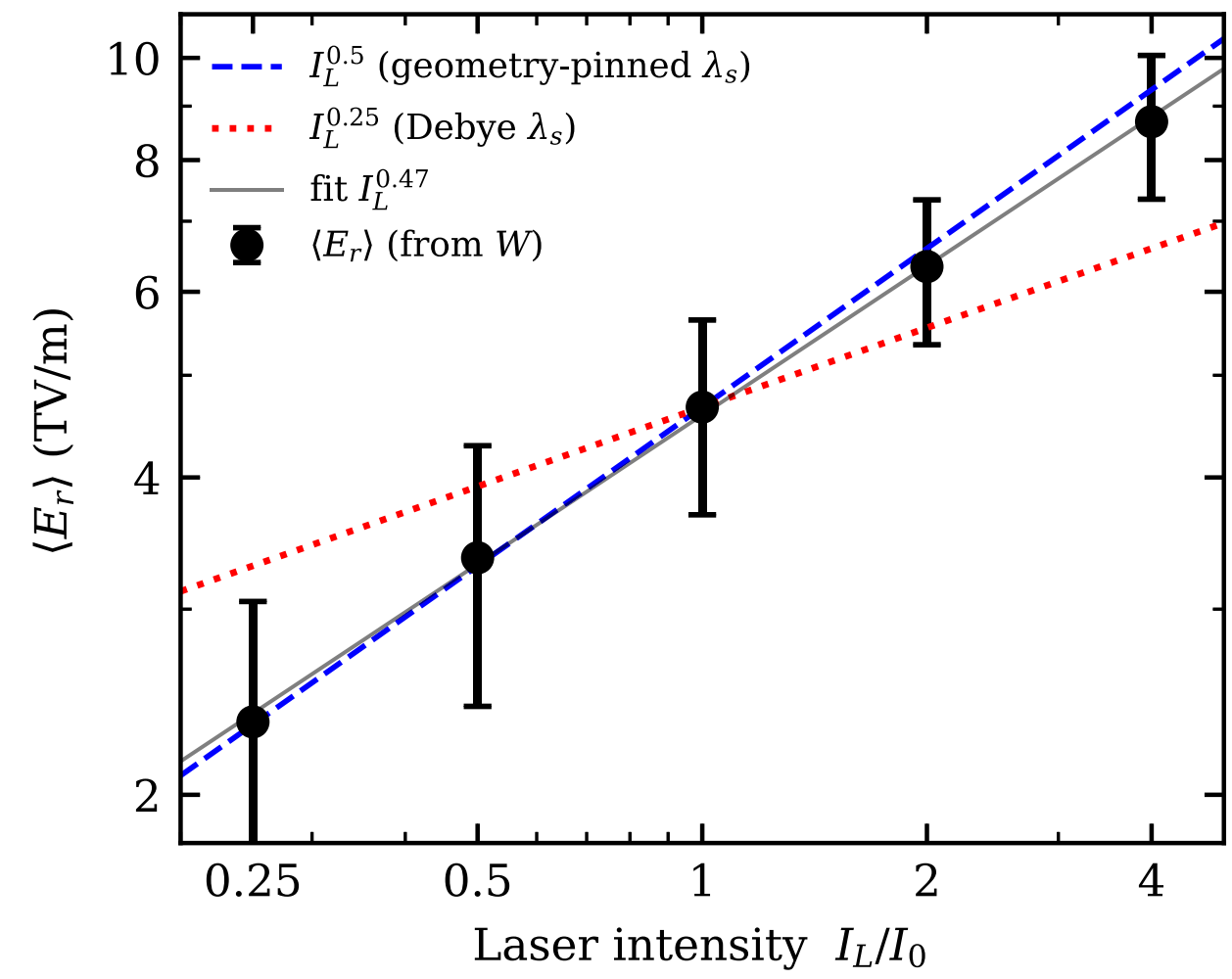

FIG. 11. Intensity scaling of the snapshot path-averaged radial field $\langle E_r \rangle_{\rm snap}$ (points: mean over four radii; error bars: standard deviation). Reference slopes denote the geometry-pinned ($I_L^{0.5}$) and Debye-limited ($I_L^{0.25}$) predictions; the measured exponent is 0.47.

The intensity scaling discriminates between two limits. If $\lambda_s$ follows the local Debye length $\lambda_D$, for which $\lambda_D \propto T_e^{1/2}$ at fixed density, then $\langle E_r \rangle_{\rm snap} \propto I_L^{0.25}$. If instead $\lambda_s$ is pinned by the cavity geometry, the field retains the ponderomotive scaling $\langle E_r \rangle_{\rm snap} \propto I_L^{0.5}$. The measured snapshot field scales as $\langle E_r \rangle_{\rm snap} \propto I_L^{0.47}$ (Fig. 11), favoring the geometry-pinned case over the explored intensity range. At $I_0$, $T_e \simeq 7$–$10$ MeV and $\langle E_r \rangle_{\rm snap} \simeq 3.1$ TV/m imply $\lambda_s \simeq 2$–$3$ μm. This scale is one to two orders of magnitude larger than the local Debye length and comparable to the convergence path.

For every intensity, $W_{\rm snap}$ is linear in $R_{\rm in}$, with correlation coefficients 0.97–0.996. At $I_0$, $dW_{\rm snap}/dR_{\rm in} = 18.8$ MeV/μm, equivalent to 3.1 TV/m and consistent with the directly measured field. The near-linearity means that the path-averaged field is approximately radius-independent over the explored range; it does not imply that the instantaneous profile itself is uniform. Thus the radius scan changes the available work mainly through the path length rather than through a systematic increase in field strength.

Motivated by this path-length dependence, we fit the intensity-averaged $\alpha$–$R_{\rm in}$ relation using a geometric-enhancement coefficient $\kappa$ and an effective radius intercept $r_0$, defined as the inner radius at which the fitted relation returns to $\alpha = 1$:

$$\alpha \simeq 1 + \kappa (R_{\rm in} - r_0). \tag{14}$$

The fit gives $\kappa = 0.53\,\mu{\rm m}^{-1}$ and $r_0 = 0.9\,\mu$m. In Eq. (12) the lower limit is fixed independently at the observed core radius $r_c^{\rm obs} \simeq 1\,\mu$m, whereas $r_0$ is an effective fit intercept. Although the two length scales are numerically close, $r_0$ may also absorb the diagnostic-shell offset and temporal gating of the moving field; no independent core-radius measurement is inferred from the fit.

The Lagrangian and snapshot diagnostics have complementary roles. The trajectory integral closes the absolute single-particle energy budget, whereas $W_{\rm snap}$ establishes the parameter dependence of the work and the effective field scale. The snapshot slope is lower than $\kappa \mathcal{E}_{\rm imp} \simeq 34$ MeV/μm at $I_0$ because the field maximum comoves with the ion front: a fixed-time profile cannot capture the full history sampled by each ion. Thus $W_{\rm snap}$ is not identified with the complete trajectory-integrated energy gain. Applied consistently across the matrix, this difference affects neither the radius linearity nor the intensity exponent.

The cancellation of intensity dependence in Eq. (14) is not exact. The measured field and incident energy scale as $I_L^{0.47}$ and $I_L^{0.37}$, respectively, so the path-work contribution has a nominal exponent difference of 0.10. Because $\alpha$ contains the additive baseline value unity, a direct power-law fit to $\alpha$ over the finite intensity range gives the smaller effective exponent $\alpha \propto I_L^{+0.067}$. In contrast, a Debye-limited field would make the path-work contribution $\alpha - 1$ decrease approximately as $I_L^{-0.12}$. The coefficient of variation of $\alpha$ across the 16-fold intensity range is 0.9%, 4.2%, 8.7%, and 13.9% for $R_{in} = 3$, 5, 7, and 10 μm. The factorization is therefore closest at small radius and remains accurate to within ∼14% over the full matrix.

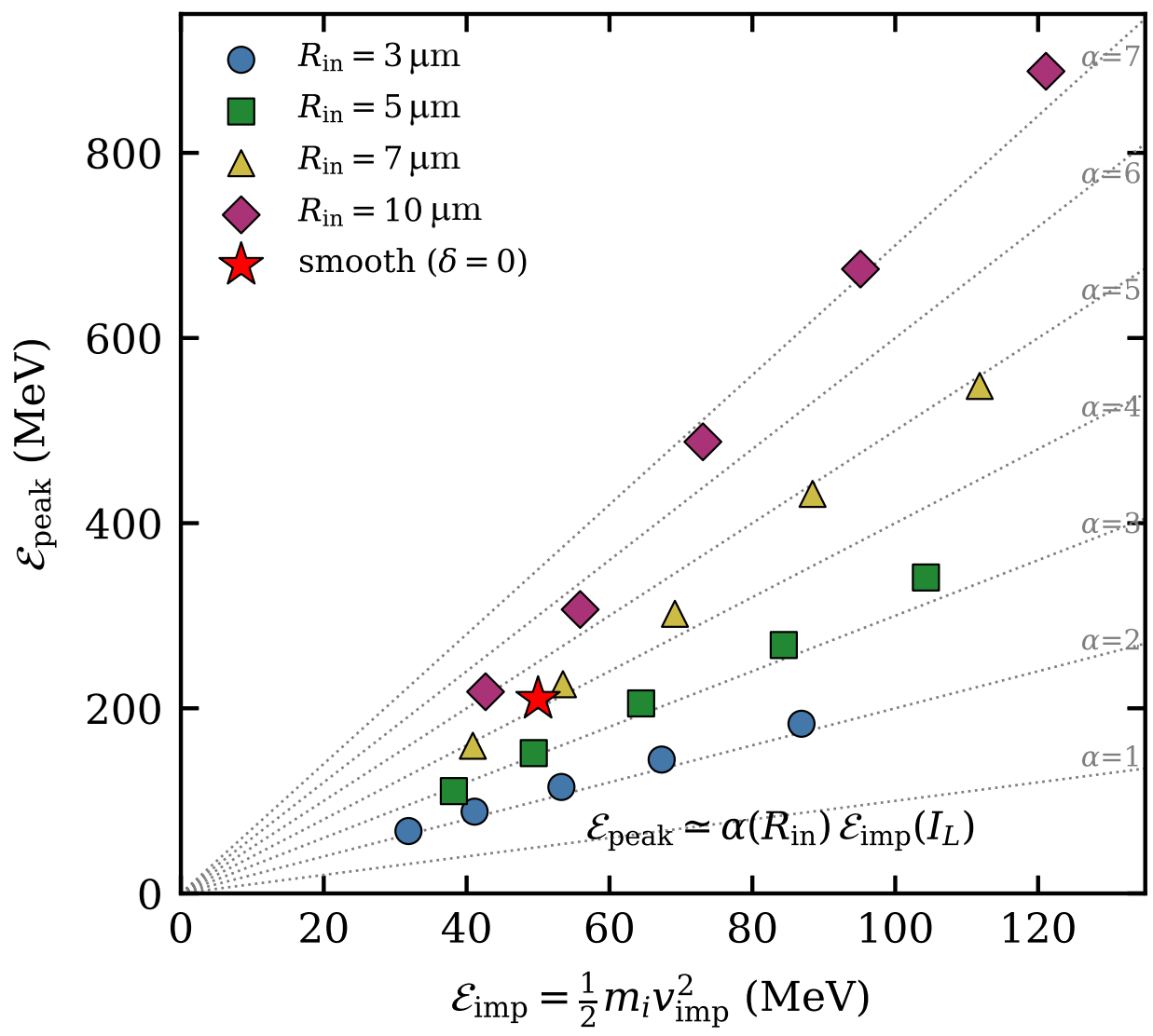

FIG. 12. $\mathcal{E}_{peak}$ versus $\mathcal{E}_{imp}$ for the 20-run matrix and the smooth-cylinder control. Dotted rays denote constant $\alpha$.

### D. Factorized scaling

The full matrix collapses onto rays of nearly constant $\alpha$ in the $(\mathcal{E}_{imp}, \mathcal{E}_{peak})$ plane (Fig. 12):

$$\mathcal{E}_{peak} \simeq \alpha(R_{in})\mathcal{E}_{imp}(I_L). \tag{15}$$

Runs with the same inner radius lie along a common ray, while changing laser intensity moves the result along that ray. The maximum local mean energy is 0.89 GeV at $R_{in} = 10$ μm and $I_L = 4I_0$. This is the temporal maximum of the mean energy in the central micrometer, not the spectral cutoff. The smooth-cylinder control ($\mathcal{E}_{peak} \simeq 210$ MeV, $\alpha \simeq 4.2$) lies in the same scaling band, supporting the conclusion that peak energization is intrinsic to cylindrical convergence.

## VI. DISCUSSION

### A. Role of blade modulation

The smooth and bladed targets reach similar peak energies (210 and 205 MeV), showing that blade modulation is not the source of peak energization. Their incident velocities differ, however. The modulated wall has a larger effective emitting area and locally enhanced sheath fields near the blade tips. This gives the bladed target a modestly larger $v_{imp}$ and therefore a lower $\alpha$ at nearly the same $\mathcal{E}_{peak}$ (3.1 for the bladed target and 4.2 for the smooth target). The smooth target also shows a slightly larger diagnostic difference between the peak local mean energy and the incident-flow energy scale, $\mathcal{E}_{peak} - \mathcal{E}_{imp} \simeq 160$ MeV for the smooth target versus 140 MeV for the bladed target, consistent with blade-induced azimuthal deflection redirecting part of the flow. These quantitative differences do not alter the central conclusion: peak energy is governed primarily by the cylindrical convergence and its radial work integral.

Blade modulation does modify the poststagnation dynamics. It increases the time-integrated number of ions above 50 MeV in $r < 1$ μm by a factor of 1.6 and doubles the central energy-time integral (Fig. 13).

In the bladed target, an organized azimuthal Fourier mode with mode number $m = 8$ develops at $r \simeq 2$–4 μm after stagnation and carries 40%–50% of the local $B_z$ spectral power. The field is several times stronger than in the smooth target at those radii. Inside $r < 1$ μm; however, the magnitude is comparable in both geometries. For fields of several hundred kT, the Larmor radius of ions below about 100 MeV is comparable to the ring radius. The structured layer can therefore deflect part of the outward-moving population and extend its central residence, without materially changing the peak energy.

The system thus has three approximately independent controls: laser intensity sets the incident-flow energy, inner radius sets the electrostatic path-length enhancement, and blade modulation sets the poststagnation magnetic organization. Transient high-$m$ structure near the ion skin-depth scale may reflect early ion-Weibel filamentation, but it is soon replaced by system-scale modes [25]. The measured growth rate of the central magnetic energy, $\sim 5 \times 10^{12}\,s^{-1}$, is also about an order of magnitude below the estimated Weibel rate. Such filamentation may contribute to early velocity-space broadening but does not appear to control the later central field or ion residence.

### B. Physical scope and limitations

The mechanism shares its sheath-field energy source with target-normal sheath acceleration [2,3], but cylindrical convergence adds a path-length enhancement absent from planar expansion. Unlike collisionless shock acceleration [6,7], no propagating density discontinuity develops, and the central distribution remains non-Maxwellian and anisotropic; unlike collisional ICF stagnation, Coulomb collisions do not establish an approximately isotropic pressure on the relevant timescale. Within the microtube literature, earlier work emphasized magnetic-field generation; the present result separates that magnetic organization from the electrostatic origin of the ion energy.

The comparison with radiation-pressure lensing [17] is instructive. There radiation pressure directly drives and deforms a solid-density target, and convergence is exploited to focus the accelerated ion layer. Here ions are launched from the inner wall and subsequently traverse the hollow cavity, where a self-consistent radial electrostatic field continues to perform work over an extended portion of the inward path. The distinctive role of geometry identified here is therefore not only to converge the flow, but also to set the interaction length over which electrostatic work can accumulate.

Microbubble implosion [26] provides a complementary limit. There, a nanometer-scale, nonquasineutral cavity releases electrostatic energy through Coulomb explosion.

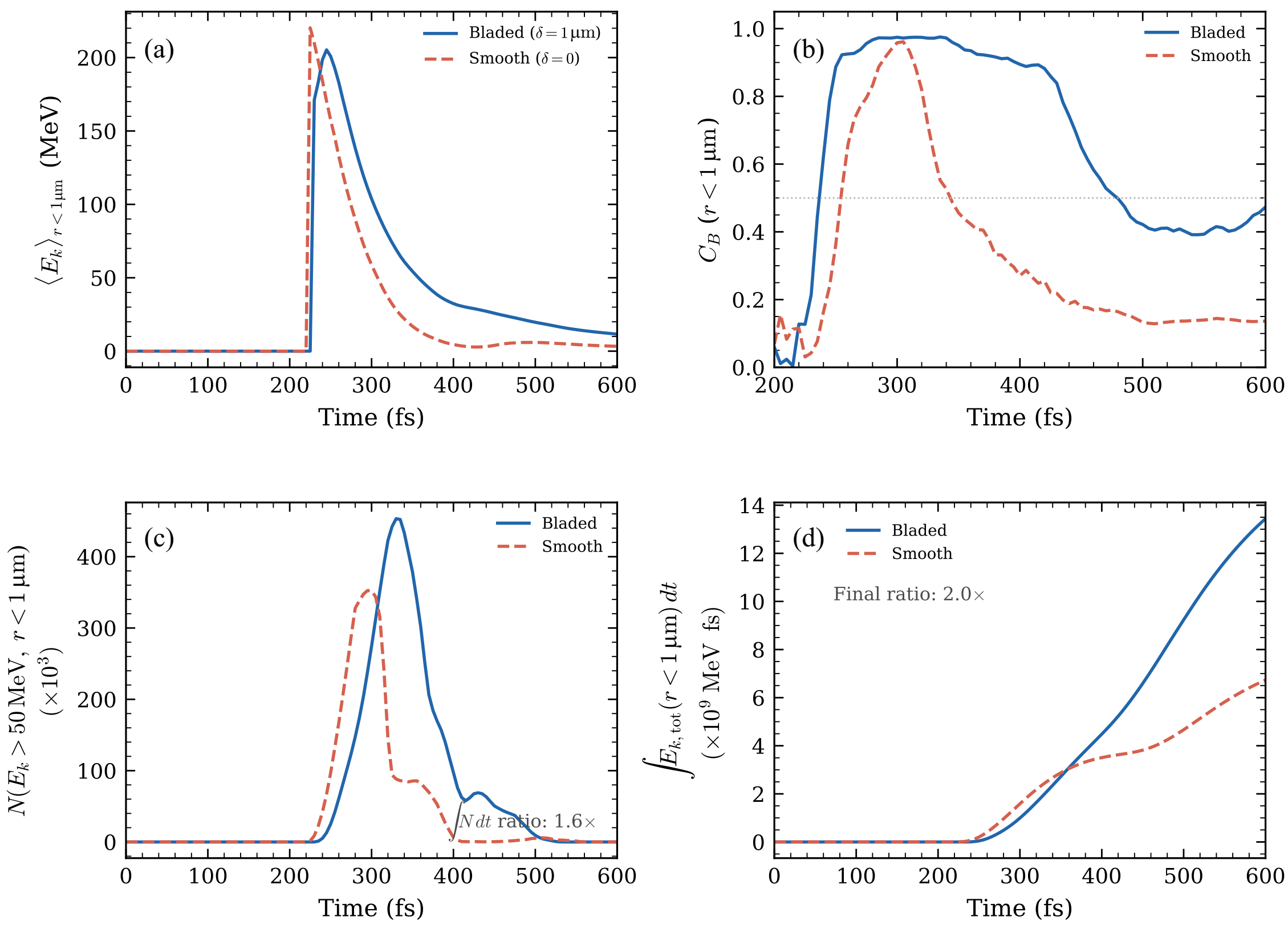

FIG. 13. Smooth and bladed targets at $R_{\rm in} = 5\,\mu$m and $I_L = I_0$: (a) central mean ion energy, (b) magnetic-field coherence $C_B \equiv |\langle B_z\rangle_A|/\langle|B_z|\rangle_A$, where $\langle\cdot\rangle_A$ denotes an area average over $r < 1\,\mu$m, (c) number of ions above 50 MeV, and (d) time integral of the total ion kinetic energy in $r < 1\,\mu$m.

The present system satisfies $R_{\rm in} \gg \lambda_D$ and remains nearly quasineutral, so the dominant energization instead arises from cumulative electrostatic work along the convergence path.

The present study identifies an energization mechanism and its scaling rather than evaluating an optimized ion source. The reported $\mathcal{E}_{\rm peak}$ is the peak local mean energy within the central micrometer during stagnation, not a spectral cutoff, and the central population is broad and non-Maxwellian. A source-oriented assessment would additionally require quantities such as the spectrum and extractable charge, downstream divergence and emittance, conversion efficiency, and a defined extraction geometry; these are outside the present scope. Accordingly, we do not assess here the utility of this configuration as an ion source for inertial-fusion applications. The comparison with ICF stagnation above is instead intended as a physical contrast between collisionless and collisional convergence.

Within the two-dimensional dynamics simulated here, the identification of the dominant energization channel is well constrained: the ensemble decomposition assigns more than 98% of the in-plane electric work to the radial channel, while the trajectory accounting reproduces individual energy gains with a regression slope of 1.01 and a median relative closure error of 8%; the radial share of the single-particle electric work has a median value of 97%. What is not established is whether this quantitative partition persists in three dimensions. Finite axial extent and axial ion escape introduce additional transport and loss channels, while three-dimensional electron transport and magnetic-field topology may also modify the self-consistent radial accelerating field. The resulting quantitative changes in $\mathcal{E}_{\rm peak}$ and poststagnation residence therefore cannot be inferred from the present two-dimensional simulations.

The ion-species dependence has not been tested. Under the assumptions of the reduced model, the normalized convergence enhancement contains no explicit leading-order dependence on $Z$ or $m_i$: retaining the explicit species dependence, $v_{\rm imp} \sim \sqrt{ZT_e/m_i}$ gives $\mathcal{E}_{\rm imp} \propto ZT_e$, while the electrostatic work contains the corresponding explicit factor $Z$ for otherwise comparable field and path conditions. This cancellation is therefore a leading-order expectation of the reduced description, not a demonstrated species universality. Species-dependent transit times, $t_{\rm arr} \propto \sqrt{m_i/Z}$ at fixed path length, together with the temporal gating identified above and self-consistent field evolution, can reintroduce species dependence. Dedicated multispecies simulations are required to test this expectation.

A first-principles prediction of the accelerating field also remains open. The reduced model writes the characteristic field as $\langle E_r\rangle \sim T_e/(e\lambda_s)$, and the observed intensity scaling favors a geometry-pinned effective scale over a Debye-limited one. However, no closure presently predicts $\lambda_s$ directly from the laser and cavity parameters. Such a closure would need to account for hot-electron transport, time-dependent charge separation, and pulse history. The remaining first-principles gap in the reduced model is therefore specifically the determination of this dynamic field scale.

## VII. CONCLUSION

Laser-driven collisionless cylindrical convergence energizes carbon ions mainly through radial electrostatic work accumulated along the inward path. This channel supplies more than 98% of the in-plane electric work and yields a peak energy that approximately separates into a laser-controlled incident-flow contribution and a radius-controlled geometric enhancement. The measured field scaling favors a geometry-pinned effective field scale over Debye-limited scaling, and the peak local mean ion energy reaches 0.89 GeV. The smooth-cylinder control shows that blade modulation is not required for the energization but organizes a magnetic layer that extends poststagnation ion residence. The convergence path, rather than inductive magnetic work or launch-surface acceleration alone, is therefore controls the convergence-associated enhancement of the peak local mean ion energy. This separation of energy input, geometric enhancement, and poststagnation residence provides a reduced framework for interpreting collisionless converging-flow systems. The predicted dependence of ion energization on inner radius and laser intensity offers a testable trend for future experiments on laser-driven converging targets.

## ACKNOWLEDGMENTS

We thank Siegfried H. Glenzer for many insightful discussions on laser-plasma interaction, ion acceleration, and the physical interpretation of the present work. This work was supported by Kansai Electric Power Company, Incorporated (KEPCO). All computations were performed using the supercomputer SQUID under the support of the D3 Center, The University of Osaka.

## DATA AVAILABILITY

There are no publicly available research data or software supporting this manuscript. Requests for further information or data should be sent to the authors.

## APPENDIX: BALLISTIC TWO-PEAKED VELOCITY DISTRIBUTION IN POLAR COORDINATES

Consider particles moving along straight trajectories at constant speed $v_0$ and impact parameter $h_0$ relative to the axis, with $t = 0$ defined at closest approach.

$$v_r = \frac{v_0^2 t}{\sqrt{v_0^2 t^2 + h_0^2}}, \qquad v_\phi = \frac{v_0 h_0}{\sqrt{v_0^2 t^2 + h_0^2}}.$$

As the particle crosses closest approach, $v_r$ changes from $-v_0$ to $+v_0$ and $v_\phi$ peaks. For a stream with uniform linear particle density $n_0$ along the trajectory, the particle number $N_p$ per unit radial velocity is

$$\frac{dN_p}{dv_r} = \frac{n_0 h_0 v_0^2}{\left(v_0^2 - v_r^2\right)^{3/2}}, \tag{A1}$$

so a two-peaked radial-velocity distribution can arise from coordinate geometry alone. Departure from this ballistic form is the more useful signature of velocity-space redistribution.